\documentclass[aps,prb,amsmath,twocolumn,showpacs]{revtex4-1}
\usepackage{graphicx,dcolumn,bm,amssymb,amsmath,color}
\usepackage[colorlinks=true,citecolor=blue,linkcolor=blue]{hyperref}
\usepackage[caption=false]{subfig}
\begin{document}

\def\cdag{\ensuremath c^{\dag}}
\def\c{\ensuremath c^{}}
\def\d{\ensuremath d^{}}
\def\dedag{\ensuremath d^{\dag}}
\def\x{\ensuremath \bm{X}}
\def\y{\ensuremath \bm{Y}}
\def\k{\ensuremath \bm{k}}
\def\q{\ensuremath \bm{q}}
\def\kp{\ensuremath \bm{k^\prime}}
\def\Q{\ensuremath \bm{Q}}
\renewcommand{\Re}{\operatorname{Re}}
\renewcommand{\Im}{\operatorname{Im}}

\title{Particle-hole condensates of higher angular momentum in hexagonal systems}

\author{Akash V. Maharaj${}^{1}$}

\author{Ronny Thomale${}^{1,2,3}$}

\author{S. Raghu${}^{1,4}$}

\affiliation{${}^{1}$Department of Physics, Stanford University, Stanford, California 94305, USA}
\affiliation{${}^{2}$Institut de th\'eorie des ph\'enom\`enes physiques, \'Ecole Polytechnique F\'ed\'erale de Lausanne (EPFL), CH-1015 Lausanne}
\affiliation{${}^3$Institute for Theoretical Physics, University of W\"urzburg, Am Hubland, D 97074 W\"urzburg}
\affiliation{${}^{4}$SLAC National Accelerator Laboratory, Menlo Park, CA 94025, USA}

\date{\today}

\begin{abstract}
{Hexagonal lattice systems ({\it e.g.} triangular, honeycomb, kagome) possess a multidimensional irreducible representation corresponding to $d_{x^2-y^2}$ and $d_{xy}$ symmetry.  Consequently, various unconventional phases that combine these $d$-wave representations can occur, and in so doing may break time-reversal and spin rotation symmetries. We show that hexagonal lattice systems with extended repulsive interactions can exhibit instabilities in the particle-hole channel to phases with either $d_{x^2-y^2}+d_{xy}$ or $d+id$ symmetry.    When lattice translational symmetry is preserved,  the phase corresponds to nematic order in the spin-channel with broken time-reversal symmetry, known as the $\beta$ phase.  On the other hand, lattice translation symmetry can be broken, resulting in various $d_{x^2-y^2}+d_{xy}$  density wave orders.  In the weak-coupling limit, when the Fermi surface lies close to a van Hove singularity, instabilities of both types are obtained in a controlled fashion.    }
\end{abstract}
\maketitle

\section{Introduction.} 
Correlated electron materials often exhibit  a tendency towards forming multiple, competing  phases including superconductivity, density waves, and phases with orientational order.  In the weak-coupling limit, the set of all possible broken symmetries  is determined by  symmetry alone; each is labeled by  an allowed irreducible representation (irrep) of the symmetric normal state.  When symmetry allows for higher dimensional irreps, unconventional phases that combine these representations are possible. Some of these can spontaneously break time-reversal and/or spin rotation symmetry. Examples include a $p_x + i p_y$ superconductor in a tetragonal system, or the $B$-phase of Helium-3 in a cubic crystal.  The former breaks time-reversal, whereas the latter represents a phase with broken spin rotation symmetry, i.e. an example of dynamically generated spin-orbit coupling.  

Here, we study systems with hexagonal symmetry where the $d_{x^2-y^2}$ and $d_{xy}$ irreps are degenerate, forming a two-component irreducible representation.~\cite{Sigrist} This allows for the possibility of $d+id$ superconductivity, which spontaneously breaks parity and time-reversal symmetry.~\cite{Hornerkamp,Sri_rg,NandkishoreSC12,RonnyGraphene12,kiesel-cm1301}   However, superconductivity is not the only prospect:  hexagonal systems can also naturally allow for {\it particle-hole} condensates which combine $d$-wave representations.  We find that when such a system preserves lattice translation symmetry, a $d+id$ particle-hole condensate known as the $\beta$ phase~\cite{ShouChengBeta07} can form: it is an electron nematic state in the spin channel that spontaneously breaks time-reversal and spin-rotation symmetry. In this phase, spin orbit coupling is generated dynamically.~\cite{wu2004dynamic} Conversely, when the particle-hole condensate breaks translation symmetry, a $
 d_{x^2-y^2} + d_{xy}$ density wave phase can form. This can occur either in the spin singlet or triplet channel,~\cite{Nayak} and accompanies a conventional charge density wave phase. $d$-density wave states were first proposed in a different context, as candidates for the enigmatic pseudogap phase in the under-doped cuprate superconductors.\cite{chakravartydensity} 

In this paper, we demonstrate the existence of both the $d_{x^2-y^2}+d_{xy}$ density wave, and the $\beta$ phase on a triangular lattice.  We show here that both phases are stablilized in the weak-coupling limit when the Fermi surface is tuned to cross a van Hove singularity.  In such a regime, there is an instability to each of these phases allowing the use of an unrestricted Hartree-Fock treatment.  Although this fine tuning is required to access the instabilities in a controlled fashion,  it is plausible that such phases survive away from the van Hove filling in the experimentally relevant intermediate coupling regime.  We also study the  transition between the two phases as the degree of Fermi surface nesting is altered.  Since both represent particle-hole condensates of higher relative angular momentum, they require longer range interactions in order to be stabilized over their zero angular momentum counterparts - namely ferromagnetism and spin density wave phases, respectively.~\cite{PhysRevB.39.2940}  Since the considerations below are based solely on symmetry, these results are also appropriate on other hexagonal lattices such as the honeycomb and kagome, as has been confirmed by recent studies.~\cite{RonnyGraphene12,wang2012frg,RonnySublattice12,RonnyKagome12,2012arXiv1210.4338T,NandkishoreSDW,NandkishoreSDWSC} 

This paper is organized as follows. In the next section we introduce the extended Hubbard model used in our analysis, and describe how instabilities corresponding to different irreps of the triangular lattice are obtained. In Sec.~\ref{sec:mf} we provide a mean field description of these phases, before obtaining exact expressions for the RPA susceptibilities in Sec.~\ref{sec:analyst}. These are used along with the Hartree-Fock variational results of Sec.~\ref{sec:hf} to discuss the various phases present in this model. In Sec.~\ref{sec:LG} we explain how Landau-Ginzburg theory justifies these mean field solutions, before discussing these results in Sec.~\ref{sec:disc}.

\section{Extended Hubbard model}\label{sec:model}
We consider spinful fermions hopping on a triangular lattice, with nearest and next nearest neighbor (nnn) hopping strengths, $t_1$ and $t_2$, and repulsive on-site and nearest neighbor density-density interaction. The Hamiltonian is
\begin{align}
H = \sum_{\bm{k},\alpha} \xi_{\bm{k}}\,c^{\dag}_{\bm{k}\alpha} c^{}_{\bm{k}\alpha} +U\sum_{i} n_{i\uparrow} n_{i\downarrow} + V\sum_{\langle i,j \rangle}\sum_{\alpha,\alpha^{\prime}} n_{i\alpha} n_{j\beta}, \label{eq:model}
\end{align}
where $n_{i\alpha} = c^{\dag}_{i\alpha}c^{}_{i\alpha}$ is the density operator for electrons of spin $\alpha$, at site $i$. We have used $\xi_{\k} = \varepsilon_{\k} -\mu$, where the dispersion is $\varepsilon_{\bm{k}} = -2t_1 \left[ \cos{k_x} + 2\cos{(k_x/2)}\cos{(\sqrt{3}k_y/2)}\right]  \quad - 2t_2\left[ \cos{\sqrt{3}k_y} + 2\cos{(3k_x/2)}\cos{(\sqrt{3}k_y/2)}\right] $. There is a logarithmic divergence in the density of states at the van Hove filling of $\mu^{*} = 2\left( t_1 + t_2 \right)$, which corresponds to $3/4$ filling when the second neighbor hopping is zero. When this is the case, the Fermi surface is perfectly nested in three inequivalent directions as shown in Fig.\ref{fig:fermisurface}. However this is not generic, and further neighbor hopping destroys the perfect nesting.

The Fourier transformed interaction can be written as
\begin{align}
&H_\text{int} = -U\sum_{\bm{k},\bm{k}^\prime,\bm{q}}  c^{\dag}_{\bm{k}\downarrow}c^{}_{\bm{k+q}\uparrow}c^{\dag}_{\bm{k^\prime+q}\uparrow}c^{}_{\bm{k^\prime}\downarrow}\nonumber \\
&- \frac{V}{2}\sum_{\bm{k},\bm{k}^\prime,\bm{q}}\sum_{\alpha,\beta,\gamma,\delta} f^{}_{\bm{k},\bm{k}^\prime} (c^{\dag}_{\bm{k}\alpha} \tau^{\nu}_{\alpha\beta}\c_{\k+\q\beta})(\cdag_{\kp+\q \gamma}\tau^{\nu}_{\gamma\delta}\c_{\kp\delta}),
\end{align}
which is attractive in the particle hole channel, with the nearest neighbor interaction suggesting instabilities in non-zero angular momentum channels. There is an implicit sum over $\nu$ here, with the identity $\tau^{0}_{\alpha\beta}=\delta_{\alpha\beta}$, and $\tau^{i}_{\alpha\beta}$ are the usual Pauli matrices in spin space. From symmetry considerations, it follows that $f^{}_{\bm{k},\bm{k}^\prime}$ can be written as a sum of separable interactions in each of the distinct irreps of the normal state point group. These can be thought of as lattice analogs of Landau parameters. For hexagonal systems with a $D_{6h}$ point group, this sum can be written as
\begin{align}
f^{}_{\bm{k},\bm{k}^\prime} &= \frac{1}{3} d^{(A_{1g})}_{\bm{k}} d^{(A_{1g})}_{\bm{k}^\prime} + \frac{2}{3} \bm{d}^{(E_{2g})}_{\bm{k}}\cdot\bm{d}^{(E_{2g})}_{\bm{k}^\prime} \nonumber \\
&\quad+ \frac{1}{3}d^{(B_{2u})}_{\bm{k}}d^{(B_{2u})}_{\bm{k}^\prime} + \frac{2}{3} \bm{d}^{(E_{1u})}_{\bm{k}}\cdot\bm{d}^{(E_{1u})}_{\bm{k}^\prime},
\label{eq:sumirrep}
\end{align}
where $d^{(A_{1g})}_{\bm{k}}$ is the explicit form of the $A_{1g}$ or extended $s$-wave representation etc. The other irreducible representations present are $f$-wave, and the two doubly degenerate representations relate to $d$-wave and $p$-wave. For the $E_{2g}$ correlations which are of interest here, the explicit $\{d_{x^2-y^2},d_{xy}\}$ form factors are $\{\cos{k_x}-\cos{\frac{k_x}{2}}\cos{\frac{\sqrt{3}k_y}{2}},\sqrt{3}\sin{\frac{k_x}{2}}\sin{\frac{\sqrt{3}k_y}{2}}\}$.

\section{Fermi surface instabilities}\label{sec:mf}
 For purely repulsive interactions, there is no superconducting instability to first-order in the interactions - {\it i.e.} at mean field level -  since there are no negative eigenvalues of the interaction matrix. We therefore examine instabilities in the particle-hole channel towards the formation of phases with order parameters $\Delta^{(\eta)}_{\bm{q};\alpha,\beta} = \sum_{\bm{k}} V^{(\eta)}d^{(\eta)}_{\bm{k}} \langle c^{\dag}_{\bm{k+q}\alpha} c^{}_{\bm{k}\beta} \rangle$, where $\eta$ labels the irrep. Having decomposed the interaction into separate irreps, we see that in the Random Phase Approximation (RPA) the effective interaction will not mix different representations. The  effective interaction in RPA, $\Gamma_{\bm{k,k^\prime}}$, will have the form
\begin{equation}
\Gamma^{}_{\bm{k,k^\prime}} = \sum_{\eta} \frac{V^{(\eta)}}{1-V^{(\eta)}\chi^{(\eta)}(\bm{q})}\, d^{(\eta)}_{\bm{k}}d^{(\eta)}_{\bm{k}^\prime},
\end{equation}
where $\chi^{(\eta)}(\bm{q})$ are the non-interacting susceptibilities, and the sum is over \textit{distinct} irreps $\eta$ as in Eq. \ref{eq:sumirrep}.
The free susceptibility for each irrep diverges logarithmically at the van Hove filling so the instabilities exist at infinitesimal bare coupling. The dominant instability is then determined by the irrep whose logarithm has the largest pre-factor.
\begin{figure}
\begin{center}
        \subfloat[{ }]{
        \includegraphics[width=0.2\textwidth]{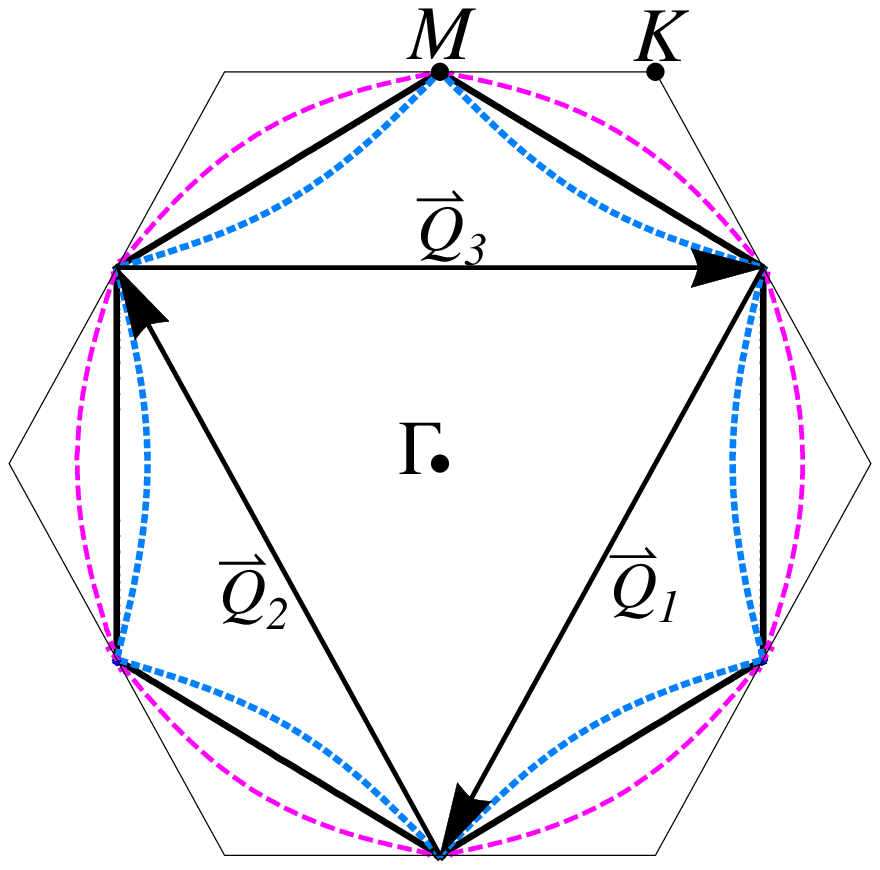}\label{fig:fermisurface}}
        \quad
        \subfloat[{}]{
        \includegraphics[width=0.2\textwidth]{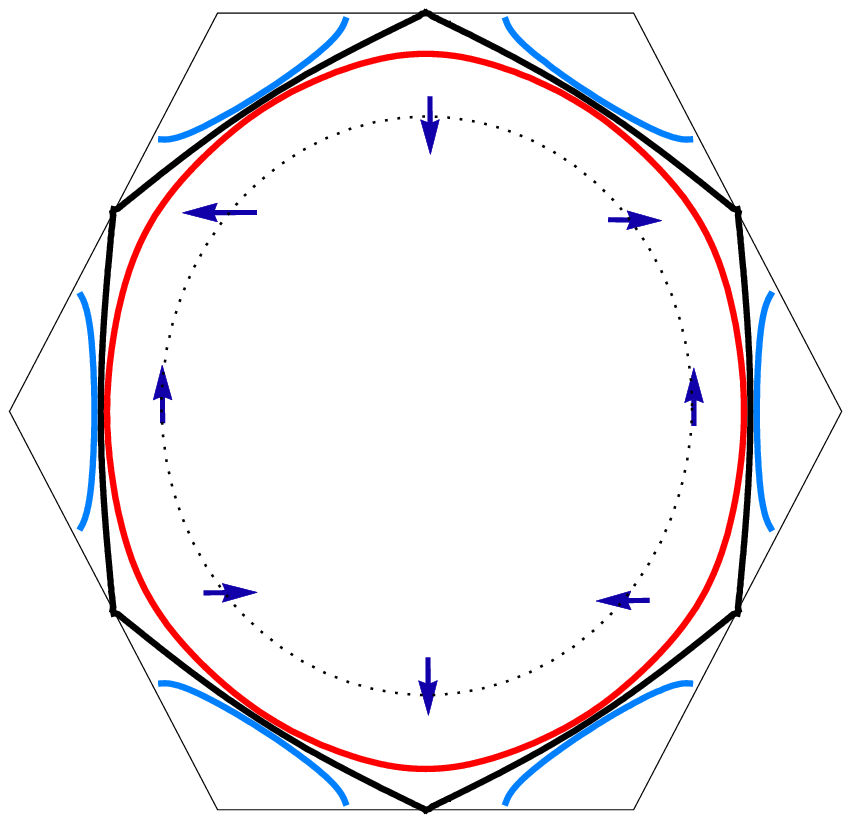}\label{fig:winding}}
       \caption{{\bf(a)}. The Fermi surface of the un-ordered state at the van Hove filling as the second neighbor hopping $t_2$ is tuned from negative (blue dashes), to zero (black) where it is perfectly nested along three vectors $\bm{Q_{1,2}}=\pi(1,\pm\sqrt{3})$ and $\bm{Q_3} = 2\pi(1,0)$,  and finally to positive (pink dashes). {\bf(b)} The Fermi surface splitting from the original black surface, into red and blue contours upon entering the $\beta$-phase; the van Hove points are removed, and the spin winds twice around the Brillouin zone. The spin orientation is opposite on each Fermi surface, and the inner circle shows winding on one of these. }
       \label{fig:lattice}
\end{center}
\end{figure}

\textit{Mean Field Hamiltonians}- When the Fermi surface is perfectly nested, $\chi(\bm{q})$ will be most divergent at the three inequivalent nesting vectors $\bm{Q}_{1,2,3}$, and we expect density wave order. This order forms in all three directions simultaneously, expanding the unit cell to four sites in real space. A trial Hamiltonian that allows for charge density waves (CDW), spin density waves (SDW) and $d$-density waves ($d$DW) in all three directions is
\begin{align}
H_{tr}&= \sum_{\k;\alpha}\xi_{\k}\cdag_{\k\alpha} \c_{\k\alpha} +\sum_{\k;\alpha,\beta} \sum_{\Q_i} \big\{ \rho_{\Q_i}\delta_{\alpha\beta}  + S^{z} _{\Q_i}\tau^{3}_{\alpha\beta} \big\} \cdag_{\k\alpha}  \c_{\k+\Q_i\beta}  \nonumber\\
&+ \sum_{\k;\alpha,\beta} \sum_{\Q_i}\left\{  \Delta^{(1)}_{\Q_i}d^{(1)}_{\k}   +  \Delta^{(2)}_{\Q_i}d^{(2)}_{\k}  \right\} \cdag_{\k\alpha}\tau^{\nu}_{\alpha\beta} \c_{\k+\Q_i\beta}+ (h.c).
\label{eq:htr1}
\end{align}
There is a sum over the three vectors ${\Q_i}$, with CDW's denoted by $\rho_{\Q_i}$, SDW's by $S^{z}_{\Q_i}$ and $d$DW's by $\Delta_{\Q_i}$. The lattice forms of the $d$-wave irreps are  $d^{(i)}_{\bm{k}}$, and we have allowed the $d$-density wave to be in either the $\nu=0$ singlet, or triplet ($\nu=1,2,3$) channels. Without loss of generality, we have chosen the SDW to be uniaxial, as was determined from a Landau-Ginzburg analysis in Ref.~\onlinecite{NandkishoreSDW}.

While we can expect density wave order when there is perfect nesting, the generic scenario of finite second neighbor hopping destroys this condition as shown in Fig.~\ref{fig:fermisurface}. We may therefore expect  $\bm{q}=\bm{0}$ (Pomeranchuk) orders to be favored beyond some critical value of $t_2$. We will see that an $\ell=2$ Pomeranchuk instability (i.e. a nematic state) becomes dominant for large enough $V$ and $t_2$. While such instabilities can occur in the charge channel, there is the more exotic possibility of realizing a $d+id$ nematic phase in the spin channel. This is favored because it utilizes both irreps, thus gaining condensation energy. This so called $\beta$ phase has a trial Hamiltonian of the form
\begin{align}
H_{tr}&= \sum_{\bm{k};\alpha,\beta} \left\{\xi_{\bm{k}}\delta_{\alpha\beta} + \Delta_{0}\left(d^{(1)}_{\bm{k} }\tau^{1}_{\alpha\beta} + d^{(2)}_{\bm{k}}\tau^{2}_{\alpha\beta}\right)\right\}\cdag_{\k\alpha}\c_{\k\beta}.
\label{eq:mfham}
\end{align}
$\Delta_{0}$ is the magnitude of the $\beta$ phase order parameter,  while $d^{(i)}_{\bm{k}}$ are the $d$-wave form factors as before. By diagonalizing this matrix, we see that the Fermi surface splits into two parts, with the spin winding twice in momentum space (see Fig. \ref{fig:winding}). This demonstrates how spin orbit coupling has been generated \textit{dynamically}.

\section{RPA Susceptibilities}\label{sec:analyst}
In order to determine the mean field phase diagram, we calculate the RPA susceptibilities for each of the phases outlined above. For a given irrep  $\eta$, the static susceptibility in the non interacting system is given by the expression
\begin{equation}
\chi^{(\eta)}_{}(\bm{q}) = - \int \frac{d^2k}{(2\pi)^2} \frac{f(\varepsilon_{\bm{k+q}})-f(\varepsilon_{\bm{k}})}{\varepsilon_{\bm{k+q}}-\varepsilon_{\bm{k}}} \, |d^{(\eta)}_{\bm{k}}|^2,
\label{eq:suscepty}
\end{equation}
where $f(\varepsilon_{\bm{k}})$ is the usual fermi function, and the integral runs over the first Brillouin zone. Since the zero temperature integral is dominated by the saddle points in the dispersion we can obtain analytic expressions for the nematic and density wave susceptibilities. We then explain how these connect to a numerical evaluation of Eq. \ref{eq:suscepty} at finite temperature.

For $\bm{q}=\bm{0}$ orders, the expression for the susceptibility becomes
\begin{equation}
\chi^{(\eta)}(\bm{q}\rightarrow \bm{0}) = \int \frac{d^2k}{(2\pi)^2}\,|d^{(\eta)}_{\bm{k}}|^2 \,\,\delta\left(\epsilon_{\bm{k}} - \mu\right).
\label{eq:zerotsus}
\end{equation}
Following the method of Ref.\onlinecite{hirsch2}, we can expand in polar coordinates around a single one of the six equivalent van Hove points. Choosing the point $(0,-2\pi/\sqrt{3})$, we set $k_x = r\cos{\theta}$ and $k_y = -2\pi/\sqrt{3}+ r\sin{\theta}$, and so, the $\bm{k}$ integral can be transformed into integrals over $\theta$ and $r$.
\begin{figure}[t]
\begin{center}
        \includegraphics[width=0.45\textwidth]{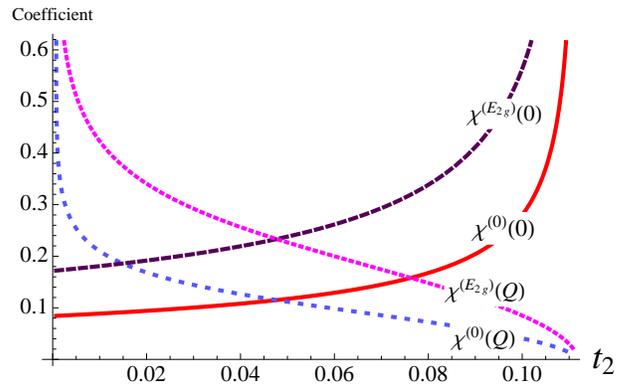}
       \caption{The pre-factors of the logarithmically diverging susceptibilities at zero temperature from Eq.~\ref{eq:dos}, \ref{eq:did}, \ref{eq:dw} and \ref{eq:diddw}. This shows crossover from density wave ($\bm{q}=\bm{Q}$) to Pomaranchuk $\bm{q}=\bm{0}$ orders as $t_2$ is varied. }
       \label{fig:coefficients}
\end{center}
\end{figure}
For the density of states (or ferromagnetic susceptibility), upon carrying out appropriate transformations on the delta functions of Eq. \ref{eq:zerotsus}, we obtain
\begin{equation}
\chi^{(0)}(\bm{0})\equiv\rho(0)\,\, \sim \,\,\frac{1}{2\pi^2}\sqrt{\frac{3}{(t_1 - 9t_2)(t_1-t_2)}}\,\, \log{\left(\frac{1}{x}\right)}, 
\label{eq:dos}
\end{equation}
where $x = \mu^{*}-\mu$. This shows the correct logarithmic divergence at the van Hove filling of $\mu = \mu^{*}$. Note the van Hove points disappear at $t_2 \ge t_1/9$ when the $M$ point is no longer a saddle, so we restrict our analysis to the range of parameter range $t_2 < t_1/9$.

On the other hand, for the $d$-wave Pomeranchuk susceptibility, the six van Hove points are no longer equivalent. When summed over the six saddle points, the extra factor in the integrand gives $|d^{(E_{2g})}_{\bm{k}}|^2 \approx 12 + \mathcal{O}(k^2)$. Thus, in the limit $\mu \rightarrow \mu^{*}$, the analytic expression for $d$-wave susceptibilities is
\begin{equation}
\chi^{(E_{2g})}(\bm{q}\rightarrow \bm{0}) \sim 2\rho(0). 
\label{eq:did}
\end{equation}
Note that this is the same for both the $d_{x^2-y^2}$ and $d_{xy}$ form factors, reflecting their symmetry protected degeneracy. 

For commensurate density wave order, the susceptibility in Eq.~\ref{eq:suscepty} can be exactly rewritten as 
\begin{equation}
\chi^{(\eta)}_{}(\bm{q}) = \int \frac{d^2k}{(2\pi)^2}\ \frac{f(\varepsilon_{\bm{k}})}{\varepsilon_{\bm{k+q}}-\varepsilon_{\bm{k}}}\,\left(|d^{(\eta)}_{\bm{k}}|^2 + |d^{(\eta)}_{\bm{k+q}}|^2\right) .\label{eq:susceptibilitydw}
\end{equation}
For the uniform density wave at any of the nesting vectors $\bm{Q}$, a similar expansion in polar coordinates around each van Hove point with a cutoff to ensure $\varepsilon_{\bm{k}} < 0 $ as dictated by the Fermi function gives the result
\begin{equation}
\chi^{(0)}_{}(\bm{Q}) \sim \frac{1}{\sqrt{3}\pi^2 (t_1 - 3t_2)} \log{\left[\frac{\tan{(\theta_0 + \frac{\pi}{3})}}{\tan{(\pi/3)}}\right]} \,\log{\left(\frac{1}{x}\right)},
\label{eq:dw}
\end{equation}
where $2\theta_0$ is the angle between arcs of the Fermi surface at the $M$ point, given by $2\cos{2\theta_0} = (t_1 + 3t_2)/(t_1 -3t_2)$.  This also has the requisite logarithmic divergence at the van Hove filling, but with a different coefficient to the density of states.

Note that the $d_{xy}$ and $d_{x^2-y^2}$ DW susceptibilities are no longer degenerate since the point group symmetry is broken by a finite ordering vector. This $E_{2g}$ susceptibility is a matrix in general, and transforms as such under the point group symmetry. However, the trace of this matrix is the same for all three nesting directions, and explicit evaluation gives  
\begin{equation}
\chi^{(E_{2g})}(\bm{Q}) = 2\, \chi^{(0)}(\bm{Q}),
\label{eq:diddw}
\end{equation}
which represents the same relation between $d$-wave and $s$-wave susceptibilities as for $\bm{q}=\bm{0}$ orders.

These analytic results at zero temperature provide valuable insight into the finite temperature values of the susceptibility.  At finite temperatures the logarithmic divergences are tamed, but the susceptibilities are still peaked strongly at the van Hove filling. A numerical evaluation shows that the trend in those peak values mirrors the trend of pre-factors in Fig.~\ref{fig:coefficients}. We can therefore determine the dominant instabilities for a given set of parameters ($t_2,V,U$) by using both the bare interaction strength for a given irrep ($V^{(\eta)}$), and the pre-factors of these diverging susceptibilities. The instability with the largest product of these factors has the highest transition temperature according to the generalized Stoner criterion.

For the extended Hubbard model we have considered here, the Stoner criteria for the SDW and CDW phases are
\begin{align}
1=U\chi^{(0)}(\Q),\,\,\text{and} \qquad 1 = (4V - U)\chi^{(0)}(\Q),
\label{eq:stoner1}
\end{align}
respectively, while for the $d$-density wave state we have
\begin{align}
1=\frac{2}{3}V\chi^{(E_{2g})}(\Q) = \frac{4}{3}V\chi^{(0)}(\Q).
\label{eq:stoner2}
\end{align}
Thus, for small nearest neighbor interactions $V$ (compared to on-site $U$), a spin density wave develops at all three nesting vectors. However for $V/U \ge 0.5$, the CDW state in all three directions becomes dominant. Despite the multidimensional $E_{2g}$ representation, the $d$DW phase does not have a higher transition temperature than the charge density wave. However it turns out that the CDW induces a $d$DW as we describe in Sec.~\ref{sec:hf}.

\begin{figure}[t]
\begin{center}
        \includegraphics[width=0.45\textwidth]{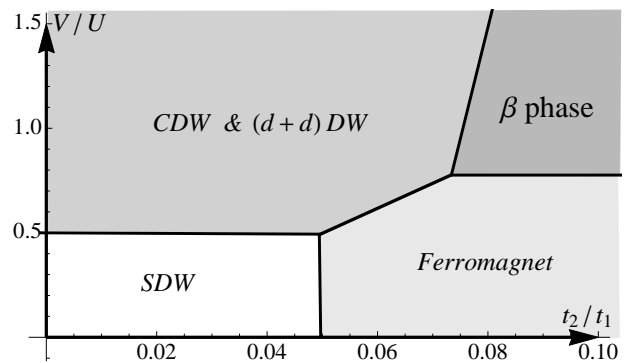}
       \caption{A schematic phase diagram of the model in Eq.~\ref{eq:model} as a function of nearest neighbor interaction $V$ and second neighbor hopping $t_2$. At low values of $V/U$, the uniform or zero angular momentum particle hole condensates dominate, while $\ell=2$ condensates are possible above a critical $V/U$. The phase boundaries, which are are first order lines at zero temperature, have been determined from the RPA results, while the combinations of phases and irreps are determined by a Hartree-Fock analysis.}
       \label{fig:phase_diag}
\end{center}
\end{figure}

For $\q=0$ only two types of particle-hole condensates are possible. The Stoner criteria for these ferromagnetic, and nematic phases respectively are
\begin{align}
1=U\rho(0),\,\,\text{and}\qquad 1=\frac{2}{3}V\chi^{(E_{2g})}(\bm{0})=\frac{4}{3}V\rho(0).
\end{align}
The result of comparing these relations with those in  Eq.~\ref{eq:stoner1} and Eq.~\ref{eq:stoner2},  is a variety of phases in the plane of $V/U$ vs. $t_2$. However, while the analytic results of this RPA analysis provide the dominant instabilities and the boundaries between them, they do not give the exact combination of order parameters in each phase. We must therefore use a Hartree-Fock variational approach to determine this combination.

\section{Hartree-Fock analysis}\label{sec:hf}
To supplement the RPA results, we numerically solve the Hartree-Fock variational equations by minimizing the free energy $F_0 = F_{tr} + \langle H - H_{tr}\rangle_{tr}$. The trial Hamiltonian is a combination of Eq.~\ref{eq:htr1} and~\ref{eq:mfham}, which allows for all possible order parameters, and all averages are taken with respect to the trial Hamiltonian. We obtain a set of self consistency equations for all order parameters, and numerically iterate these equations until convergence is achieved. 

The result of this analysis along with the RPA results, is the schematic phase diagram of Fig.~\ref{fig:phase_diag}. For small $V/U$, and depending on the degree of nesting present, the conventional SDW and Ferromagnetic (i.e. $s$-wave or $\ell=0$ orders) are stable. However, higher angular momentum $\ell=2$ particle hole condensates are possible when longer range interactions are significant. There is however, a distinction between phases which preserve and break translation symmetry. When translation symmetry is broken, there is a real superposition of $d$DW's of the form $d_{x^2-y^2} + d_{xy}$, which coexist with the charge density wave. The CDW has equal magnitude for all three ordering vectors, while the $d$DW components transform as a vector. There is a first order transition into this mixed phase as temperature is lowered.

Conversely, when translation symmetry is preserved with $V$ significant compared to $U$, the two irreps acquire a phase difference of $\pi/2$ thus breaking time-reversal and spin rotation symmetry and forming the $\beta$ phase. This transition is continuous as temperature is lowered. To provide further insight into these results, we can turn to a Landau-Ginzburg analysis.

\section{Landau-Ginzburg theory}\label{sec:LG}
In order to understand why the nematic phase is $d+id$, but the $d$DW contains a real superposition of irreps, we must look at the Landau-Ginzburg ($LG$) free energy expansions up to fourth order.

Starting with the nematic phase, our numerical calculations confirm that these transitions are continuous as a function of temperature, so we can safely truncate a $LG$ free energy expansion at low order. The presence of a two component (vector) order parameter along with the $D_{6h}$ symmetry, dictates that the expansion for the translation symmetry preserving order parameters must be of the form,~\cite{Sigrist}
\begin{align}
\mathcal{F} =&\, a\left(T-T_c\right)\left(|\Delta_1|^2 + |\Delta_2|^2\right) + b_1\left(|\Delta_1|^2 + |\Delta_2|^2\right)^2 \nonumber \\
&+ b_2 \left(\Delta^{*}_1\Delta_2 - \Delta_1\Delta^{*}_2 \right)^2,
\end{align}
up to fourth order, where $\Delta_1$ and $\Delta_2$ are the $d_{x^2-y^2}$ and $d_{xy}$ components of the order parameter. The sign of the coefficient $b_2$ determines which combination of the order parameters is present; for  $b_2>0$, the sum $\Delta_1 \pm i\Delta_2$ minimizes the free energy, whereas for $b_2 < 0$  either a $d_{xy}$ or $d_{x^2-y^2}$ order results. These coefficients can be determined microscopically,~\cite{fernandes,andrey_new} and in a manner exactly analogous with superconductivity,~\cite{NandkishoreSC12} we get $b_1 = \frac{1}{2}K$ and $b_2 = \frac{1}{12}K$ where $K>0$ is a trace over four Green functions.  Thus, a $d+id$ Pomeranchuk state, i.e. the $\beta$-phase is favored at the transition.

On the other hand, when translation symmetry is broken, with CDW and $d$DW order parameters in three directions the $LG$ free energy expansion is more complicated. We can separate the expansion into three parts which involve pure CDW, pure $d$DW and coupling terms:
\begin{align}
\mathcal{F}= \mathcal{F}_{\rho} +\mathcal{F}_{\Delta} +\mathcal{F}_{\rho\Delta}.
\end{align}
Since there are order parameters in three directions, with $\Q_1+\Q_2 +\Q_3=0$, and twice any vector is the same as a reciprocal lattice vector i.e. ($2\Q_i = \bm{G}$), we have
\begin{align}
\mathcal{F}_{\rho} &= \sum^{3}_{i=1}\left(\alpha_1 \rho_{\Q_i}^2 + \beta_1\rho_{\Q_i}^4\right) + \gamma_1 \left(\rho_{\Q_1}\rho_{\Q_2}\rho_{\Q_3}\right)  \nonumber\\
&+ \sum^{}_{i\ne j}\delta_1\rho^{2}_{\Q_i}\rho^{2}_{\Q_j} .
\end{align}
Note that in general we expect the (real) order parameter to have the same magnitude\textemdash though not necessarily the same sign\textemdash in all three directions, so that both fourth order terms above can be written as one. Because we have three ordering directions, a cubic invariant is allowed in the expansion resulting in a first order transition into the CDW phase; this is indeed observed in the numerical Hartree-Fock results. Since the $d$DW has a multidimensional representation, its order parameter is a vector and no cubic term is allowed. The invariant free energy up to fourth order is therefore
\begin{align}
\mathcal{F}_{\Delta}&= \sum^{3}_{i=1}\alpha_2 \lvert \vec{\Delta}_{\Q_i} \rvert^2 +\beta_2 \lvert\vec{\Delta}_{\Q_i}\rvert^4 +\gamma_2\lvert  \vec{\Delta}_{\Q_i}\times\vec{\Delta}^{*}_{\Q_i}\rvert^2 \nonumber \\
&+\,\,\sum_{i\ne j} \beta_3 \lvert  \vec{\Delta}_{\Q_i}\cdot\vec{\Delta}^{*}_{\Q_j}\rvert^2+\gamma_3\lvert  \vec{\Delta}_{\Q_i}\times\vec{\Delta}^{*}_{\Q_j}\rvert^2 
\end{align}
As with the nematic case, the sign of $\gamma_2$ determines whether there is a real or imaginary superposition of $d$DW components at each ordering vector. In the absence of a CDW, this coefficient is positive so a $d+id$ density wave results; this is exactly analogous to the $\q=\bm{0}$ orders. However, a non-zero CDW couples to a $d$DW bilinear in the form
\begin{align}
\mathcal{F}_{\rho\Delta} &=  \lambda_{1} \rho_{\Q_1} \left( \vec{\Delta}_{\Q_2} \cdot \vec{\Delta}_{\Q_3} \right) + (\text{cyclic perms}), 
\end{align}
at lowest order. There are several higher order coupling terms which we have omitted. From the RPA susceptibilities and Hartree-Fock minimization, the CDW has a higher transition temperature than the $d$DW. Thus the quadratic $d$DW term becomes $\sim(\alpha_{2} +\lambda_{1} \rho+\ldots)|\Delta|^2$, where the dots represent higher order terms in $\rho$. Provided $\rho$ is negative and sufficiently large, the coefficient of this quadratic term will immediately become negative, allowing the $d$DW to form. By a similar argument, there is a coupling of the form $\rho^2\lvert  \vec{\Delta}\times\vec{\Delta}^{*}\rvert^2$ which changes the sign of $\gamma_2$, and results in a real superposition $d$DW's. We therefore see that these simple symmetry arguments justify the coexistence of a CDW and the real $d_{x^2-y^2} + d_{xy}$ DW.

\section{Discussion}\label{sec:disc}
We have shown on analytical footing how higher angular momentum particle hole condensates can be realized from  models with extended interactions on hexagonally symmetric lattices. The symmetry protected degeneracy of the $d$-wave representations means that 
unconventional phases which combine both irreps can form, in a manner that is analogous to superconductivity.

The full phase diagram in the plane $V$ vs. $t_2$ is shown in Fig. \ref{fig:phase_diag}. This provides a unified view of the effects of longer range interactions as bandstructure effects are tuned. When the Fermi surface is approximately perfectly nested, commensurate density wave order is dominant. For low values of $V$, a uniaxial spin density wave forms in all three nesting vectors simultaneously.~\cite{ChernGW12,NandkishoreSDW} Meanwhile, when longer range interactions dominate, a simple CDW phase emerges. However this immediately induces the higher angular momentum $d_{x^2-y^2}+d_{xy}$ density wave. This can be either a singlet or triplet density wave; the singlet state corresponds to a modulation of hopping strengths (i.e. bond order), while the triplet states are accompanied by a Goldstone boson due to the broken spin rotation symmetry. 

However, as the perfect nesting condition is weakened by finite $t_2$, the $\bm{q}=\bm{0}$ orders are favored, with a crossover from $\ell=0$ ferromagnetism to an $\ell=2$  chiral nematic phase as the interaction strength is increased. The latter is the $\beta$ phase - a $d+id$ Pomeranchuk instability which occurs in the spin channel. This is a metal with dynamically driven spin orbit coupling, i.e. one that spontaneously breaks time reversal and $SU(2)$ spin rotation symmetries. 

Such Pomeranchuk phases have been observed as same sub-lattice instabilities in recent FRG studies of Kagome lattices when very large extended interactions are present.~\cite{RonnyKagome12} While these results suggest that the $\beta$ phase may be present at intermediate coupling, we caution that results of perturbation theory can be unreliable when extrapolated to larger interactions. We also note that while at mean field level we find no instability towards superconductivity at the van Hove filling, other studies have found the singlet pairing channel to be competitive as nesting is reduced upon moving away from the saddle point.~\cite{Hornerkamp03}

While we have considered here the particular case of hexagonal systems, it is likely that non-trivial particle hole condensates can occur in cubic lattices which allow for three dimensional $t_{2g}$ representations.  It is likely, therefore, that the $\beta$ phase, which makes use of each of the irreps in the spin channel will be the favored ground state at a Pomeranchuk instability.  We shall present the analysis of three dimensional systems elsewhere. 

We acknowledge insightful discussions with Andrey Chubukov and Shou-Cheng Zhang. RT thanks M. Kiesel and C. Platt for collaborations on related topics. This work was supported in part by the U.S. Department of Energy under Contract No. AC02-76SF00515 (AM,SR), by the European Research commission through ERC-StG-2013-336012 (RT), and the Alfred P. Sloan Foundation (SR).


\begin{thebibliography}{10}

\bibitem{Sigrist}
M. Sigrist and K. Ueda, Rev. Mod. Phys. {\bf 63},  239  (1991).

\bibitem{Hornerkamp}
C. Honerkamp, Phys. Rev. Lett. {\bf 100},  146404  (2008).

\bibitem{Sri_rg}
S. Raghu, S.~A. Kivelson, and D.~J. Scalapino, Phys. Rev. B {\bf 81},  224505
  (2010).

\bibitem{NandkishoreSC12}
R. {Nandkishore}, L.~S. {Levitov}, and A.~V. {Chubukov}, Nature Physics {\bf
  8},  158  (2012).

\bibitem{RonnyGraphene12}
M.~L. Kiesel, C. Platt, W. Hanke, D.~A. Abanin, and R. Thomale, Phys. Rev. B
  {\bf 86},  020507  (2012).

\bibitem{kiesel-cm1301}
M.~L. Kiesel, C. Platt, W. Hanke, and R. Thomale, Phys. Rev. Lett. {\bf 111},
  097001  (2013).

\bibitem{ShouChengBeta07}
C. Wu, K. Sun, E. Fradkin, and S.-C. Zhang, Phys. Rev. B {\bf 75},  115103
  (2007).

\bibitem{wu2004dynamic}
C. Wu and S.-C. Zhang, Physical review letters {\bf 93},  36403  (2004).

\bibitem{Nayak}
C. Nayak, Physical Review B {\bf 62},  4880  (2000).

\bibitem{chakravartydensity}
S. Chakravarty, R. Laughlin, D. Morr, and C. Nayak, Physical Review B {\bf 63},
   094503  (2001).

\bibitem{PhysRevB.39.2940}
H.~J. Schulz, Phys. Rev. B {\bf 39},  2940  (1989).

\bibitem{wang2012frg}
W. Wang, Y. Xiang, Q. Wang, F. Wang, F. Yang, and D. Lee, Physical Review B
  {\bf 85},  035414  (2012).

\bibitem{RonnySublattice12}
M.~L. Kiesel and R. Thomale, Phys. Rev. B {\bf 86},  121105  (2012).

\bibitem{RonnyKagome12}
M.~L. Kiesel, C. Platt, and R. Thomale, Phys. Rev. Lett. {\bf 110},  126405
  (2013).

\bibitem{2012arXiv1210.4338T}
O. {Tieleman}, O. {Dutta}, M. {Lewenstein}, and A. {Eckardt},  Phys. Rev. Lett. {\bf 110},  096405
  (2013).

\bibitem{NandkishoreSDW}
R. Nandkishore, G.-W. Chern, and A.~V. Chubukov, Phys. Rev. Lett. {\bf 108},
  227204  (2012).

\bibitem{NandkishoreSDWSC}
R. {Nandkishore} and A.~V. {Chubukov}, \prb {\bf 86},  115426  (2012).

\bibitem{hirsch2}
H.~Q. Lin and J.~E. Hirsch, Phys. Rev. B {\bf 35},  3359  (1987).

\bibitem{fernandes}
R. Fernandes and J. Schmalian, Physical Review B {\bf 82},  014521  (2010).

\bibitem{andrey_new}
A.~B. Vorontsov, M.~G. Vavilov, and A.~V. Chubukov, Phys. Rev. B {\bf 81},
  174538  (2010).

\bibitem{ChernGW12}
G.-W. Chern, R.~M. Fernandes, R. Nandkishore, and A.~V. Chubukov, Phys. Rev. B
  {\bf 86},  115443  (2012).

\bibitem{Hornerkamp03}
C. Honerkamp, Phys. Rev. B {\bf 68}, 104510 (2003).


\end{thebibliography}

\end{document}